\documentclass[runningheads]{llncs}
\usepackage{graphicx}
\usepackage{amsmath}
\usepackage{algorithm}
\usepackage[noend]{algpseudocode}
\usepackage{mathtools}
\usepackage{multirow}

\usepackage{xspace}
\usepackage{amssymb}
\usepackage{acro}
\usepackage{booktabs}
\usepackage{makecell}

\newcommand{\ie}{\emph{i.e.}, }
\newcommand{\eg}{\emph{e.g.}, }

\begin{document}

\title{Semi-Supervised Learning for Bone Mineral Density Estimation in Hip X-ray Images}
\titlerunning{SSL for Bone Mineral Density Estimation in Hip X-rays}
%
%
\author{Kang Zheng\inst{1} \and
Yirui Wang\inst{1}  \and
Xiao-Yun Zhou\inst{1} \and
Fakai Wang\inst{1,2} \and
Le Lu\inst{1} \and \\
Chihung Lin\inst{3} \and
Lingyun Huang\inst{4}  \and
Guotong Xie\inst{4} \and
Jing Xiao\inst{4} \and
Chang-Fu Kuo\inst{3} \and
Shun Miao\inst{1}}
\authorrunning{Zheng et al.}
%
\institute{PAII Inc., Bethesda, MD, USA \and
University of Maryland, College Park, MD, USA \and
Chang Gung Memorial Hospital, Linkou, Taiwan, ROC \and
Ping An Technology, Shenzhen, China}

\maketitle              
\begin{abstract}

Bone mineral density (BMD) is a clinically critical indicator of osteoporosis, usually measured by dual-energy X-ray absorptiometry (DEXA). Due to the limited accessibility of DEXA machines and examinations, osteoporosis is often under-diagnosed and under-treated, leading to increased fragility fracture risks. Thus it is highly desirable to obtain BMDs with alternative cost-effective and more accessible medical imaging examinations such as X-ray plain films. In this work, we formulate the BMD estimation from plain hip X-ray images as a regression problem. Specifically, we propose a new semi-supervised self-training algorithm to train the BMD regression model using images coupled with DEXA measured BMDs and unlabeled images with pseudo BMDs. Pseudo BMDs are generated and refined iteratively for unlabeled images during self-training. We also present a novel adaptive triplet loss to improve the model's regression accuracy. On an in-house dataset of 1,090 images (819 unique patients), our BMD estimation method achieves a high Pearson correlation coefficient of 0.8805 to ground-truth BMDs. It offers good feasibility to use the more accessible and cheaper X-ray imaging for opportunistic osteoporosis screening.

\end{abstract}
\section{Introduction}

Osteoporosis is a common skeletal disorder characterized by decreased bone mineral density (BMD) and bone strength deterioration, leading to an increased risk of fragility fracture. All types of fragility fractures affect the elderly with multiple morbidities, reduced life quality, increased dependence, and mortality. FRAX is a clinical tool for assessing bone fracture risks by integrating clinical risk factors and BMD. While some clinical risk factors such as age, gender, and body mass index can be obtained from electronic medical records, the current gold standard to measure BMD is dual-energy X-ray absorptiometry (DEXA). However, due to the limited availability of DEXA devices, especially in developing countries, osteoporosis is often under-diagnosed and under-treated. Therefore, alternative lower-cost BMD evaluation protocols and methods using more accessible medical imaging examinations, \eg X-ray plain films, are highly desirable.

Previous methods aiming to use imaging obtained for other indications to estimate BMD or classify osteoporosis have been proposed. The clinical value of this ``opportunistic screening'' approach seems apparent with low additional cost, high speed to obtain essential patient information, and potentially greater accuracy over traditional prediction models. For example, previous studies~\cite{alacreu2017opportunistic,lee2017predicting,mookiah2018feasibility,jang2019opportunistic,pan2020automatic,dagan2020automated} have attempted to estimate BMD through CT scans for osteoporosis screening. However, the undesired prediction performance, high cost, long acquisition time, and particularly high radiation dose of CT imaging compared to DEXA and plain film X-ray are barriers to their wide clinical adoption. In contrast, plain film X-ray is a more accessible and lower-cost imaging tool than CT for opportunistic screening. In this work, we attempt to estimate BMD from plain film hip X-ray images for osteoporosis screening.

Our work is based on the assumption that hip X-ray images contain sufficient information on visual cues for BMD estimation. We use a convolutional neural network (CNN) for regressing BMD from hip X-ray images. Paired hip X-ray image and DEXA measured BMD taken within six months apart (as ground-truth) are collected as labeled data for supervised regression learning. To improve regression accuracy, we also propose a novel adaptive triplet loss, enabling the model to better discriminate samples with dissimilar BMDs in the feature space. Since it is often difficult to obtain a large amount of hip X-ray images paired with DEXA measured BMDs, we propose to use semi-supervised learning to exploit the usefulness of large-scale hip X-ray images without ground-truth BMDs, which can be efficiently collected at scale. Our method is evaluated on an in-house dataset of 1,090 images from 819 patients and achieves a Pearson correlation coefficient of 0.8805 to ground-truth BMDs.

Our main contributions are three-fold: 1) This work is the first to estimate bone mineral density from the hip plain films, which offers a potential computer-aided diagnosis (CAD) application to opportunistic osteoporosis screening with more accessibility and reduced costs. 2) We propose a novel adaptive triplet loss to improve the model's regression accuracy, applicable to other regression tasks as well. 3) We present a new semi-supervised self-training algorithm to boost the BMD estimation accuracy via exploiting unlabeled hip X-ray images.

\section{Method}
Given a hip X-ray image, we first crop a region-of-interest (ROI) around the femoral neck\footnote{This automated ROI localization module can be achieved by re-implementing the deep adaptive graph (DAG) network~\cite{li2020structured}.} as input to the convolutional neural network. Denoting the hip ROI as $I$ and its associated ground-truth (GT) BMD as $y$, our goal is to provide an estimation $y'$ close to $y$. We formulate the problem as a regression problem since BMD values are continuous.

\begin{figure} [t]
    \centering
    \includegraphics[width=\linewidth]{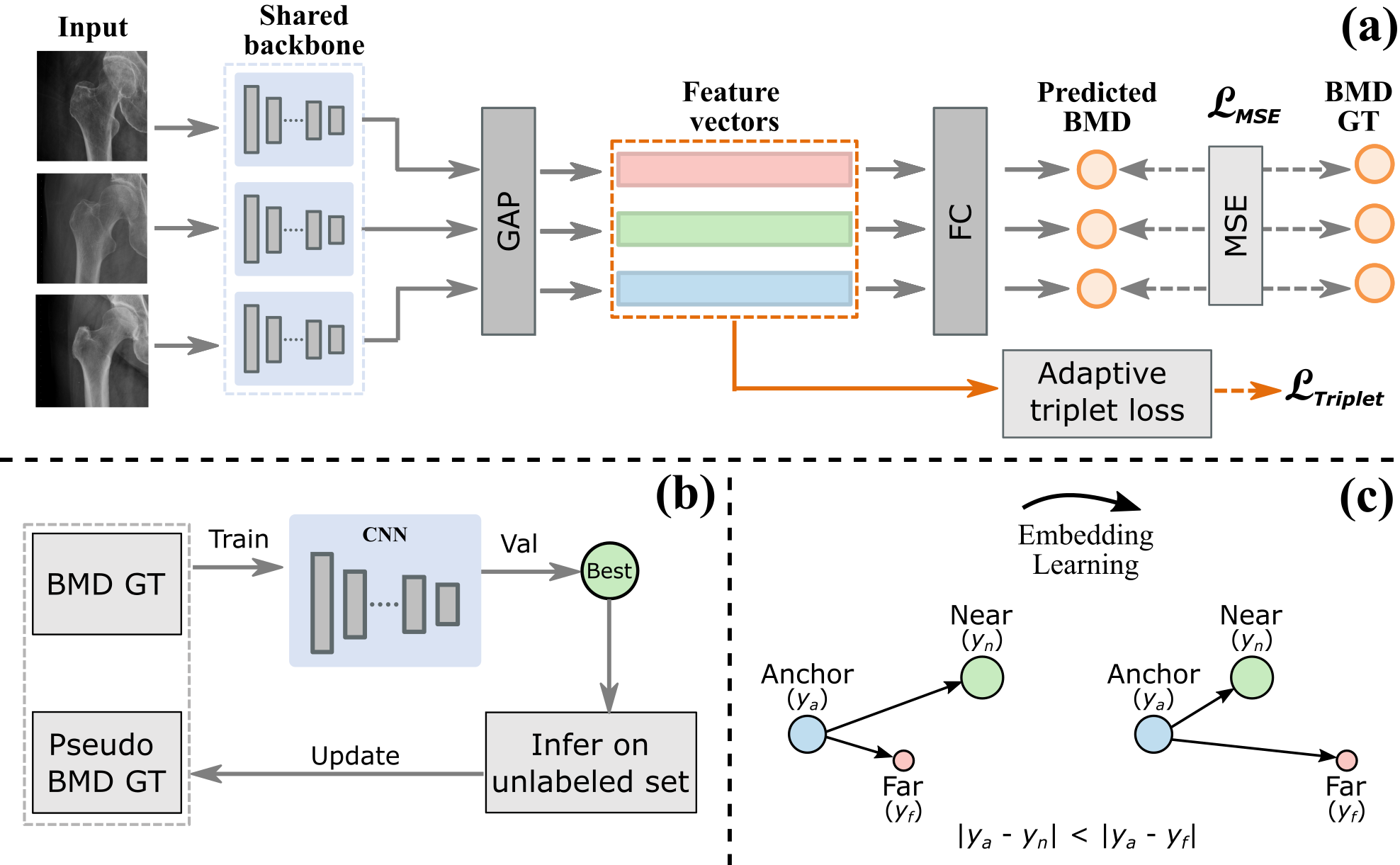}
    \caption{An overview of our proposed framework, consisting of (a) supervised pre-training stage and (b) semi-supervised self-training stage. In the supervised pre-training stage, we train the model on labeled images using the MSE loss and (c) a novel adaptive triplet loss, which encourages the distance between feature embeddings of samples correlated to their BMD difference. In the self-training stage, we fine-tune the model on labeled data and pseudo-labeled data. We update the pseudo labels when the model achieves higher performance on the validation set.}
    \label{fig:framework}
\end{figure}

Figure~\ref{fig:framework} illustrates the overall framework of our proposed method, which consists of two stages: supervised pre-training and semi-supervised self-training. In the first stage, we pre-train a model on a small amount of labeled images. We adopt VGG-11~\cite{simonyan2014very} as the backbone with its last two fully-connected (FC) layers replaced by a global average pooling (GAP) layer to obtain a 512-dimensional embedding feature vector. Then, an FC layer is applied to regress BMD from the feature vector. We train the model using Mean Square Error (MSE) loss and a novel adaptive triplet loss, as shown in Figure~\ref{fig:framework}(a). The MSE loss is defined between the predicted BMD $y'$ and the GT BMD $y$, written as:
\begin{equation}
    \mathcal{L}_{mse}=\left( y - y' \right)^2 .
\end{equation}
With the proposed adaptive triplet loss, the model learns more discriminative feature embeddings for images with different BMDs, thus improving its regression accuracy. We elaborate on the adaptive triplet loss in Section~\ref{sec:ATL}. In the second stage, we fine-tune the model with a novel semi-supervised self-training algorithm using both labeled and unlabeled images. We iteratively predict the pseudo BMDs for unlabeled images and combine them with labeled ones to fine-tune the model. The self-training algorithm is detailed in Section~\ref{sec:self_training}.

\subsection{Adaptive Triplet Loss} \label{sec:ATL}
As BMD is a continuous value, hip ROIs' embeddings are expected to be continuous in the feature space as well. In particular, the distance between embeddings of two samples in the feature space should be correlated with their BMD discrepancy. Motivated by this, we propose a novel adaptive triplet loss encouraging the model to discriminate samples with different BMDs in feature space, as shown in Figure~\ref{fig:framework}(c). For a given triplet, one sample is selected as \textit{anchor}. Among the other two samples, the one with a BMD closer to the \textit{anchor} is regarded as \textit{near} sample, and the other as \textit{far} sample. We want the embedding of the \textit{anchor} to be closer to \textit{near} sample than \textit{far} sample in the feature space by a margin, \ie
\begin{equation}
    ||F_a - F_n||^2_2 + m < ||F_a - F_f||^2_2,
\end{equation}
where $m$ is the margin. $F_a$, $F_n$, and $F_f$ are embeddings of the \textit{anchor}, \textit{near}, and \textit{far} samples, respectively. The margin that separates the {\it near} sample from the {\it far} sample should account for their relative BMD differences. Therefore, we define the adaptive triplet loss as follows:
\begin{equation}
    \mathcal{L}_{triplet}=\left[ ||F_a - F_n||^2_2 - ||F_a - F_f||^2_2 + \alpha m \right]_{+}.
\end{equation}
$\alpha$ is the adaptive coefficient based on the BMD differences, defined as:
\begin{equation}
    \alpha = ||y_a - y_f||^2_2 - ||y_a - y_n||^2_2,
    \label{eq:adaptive_coefficient}
\end{equation}
where $y_a$, $y_n$, and $y_f$ are the GT BMD values of the {\it anchor}, {\it near}, and {\it far} samples, respectively. For network training, we combine the MSE loss with adaptive triplet loss as follows:
\begin{equation}
    \mathcal{L}=\mathcal{L}_{mse}+\lambda \mathcal{L}_{triplet},
\end{equation}
where $\lambda$ is the weight for adaptive triplet loss. We set $\lambda$ to 0.5 for all experiments.

\subsection{Semi-Supervised Self-Training} \label{sec:self_training}


Given limited images coupled with GT BMDs, the model can easily overfit the training data and yield poor performance on unseen test data. To overcome this issue, we propose a semi-supervised self-training algorithm to leverage both labeled and unlabeled data. An overview of the proposed self-training algorithm is illustrated in Figure~\ref{fig:framework}(b). 
We first use the pre-trained model to predict pseudo GT BMDs on unlabeled images to obtain additional supervisions. The unlabeled images with pseudo GT BMDs are subsequently combined with labeled images to fine-tune the model. To improve the quality of estimated pseudo GT BMDs, we propose to refine them using the better fine-tuned models during self-training. We assume that if a fine-tuned model achieves higher performance on the validation set, it can also produce more accurate/reliable pseudo GT BMDs for unlabeled images. Specifically, after each self-training epoch, we evaluate the model performance on the validation set using Pearson correlation coefficient and mean square error. If the model indeed achieves higher correlation coefficient and lower mean square error than previous models at the same time, we use it to re-generate pseudo GT BMDs for the unlabeled images for self-training. This process is repeated until the total self-training epoch is reached. We summarize this semi-supervised self-training algorithm in Algorithm~\ref{alg:self_training}.

In addition, we augment each image twice and employ consistency constraints between their features and between predicted BMDs, which regularize the model to avoid being misled by inaccurate pseudo labels. Denoting two augmentations of the same image as $I_1$ and $I_2$, their features $F_1$ and $F_2$, and their predicted BMDs $y_1$ and $y_2$, we define the consistency loss as:
\begin{equation}
    \mathcal{L}_{c} = ||F_1 - F_2||^2_2 + ||y_1 - y_2||^2_2.
    \label{eq:consistency_loss}
\end{equation}
For self-training, the total loss is:
\begin{equation}
    \mathcal{L}=\mathcal{L}_{mse}+\lambda \mathcal{L}_{triplet} + \lambda' \mathcal{L}_{c},
\end{equation}
where $\lambda'$ is the consistency loss weight set to 1.0.

\begin{algorithm} [t]
\caption{Semi-supervised self-training algorithm}\label{alg:self_training}
\begin{algorithmic}[1]
\State Initialize the best R-value $\hat{\eta}\coloneqq0$ and MSE $\hat{\epsilon}\coloneqq \infty$
\State Initialize training epoch $e\coloneqq0$ and set total training epoch $E$
\State Initialize model with pre-trained weights
\While{$e<E$}
 \State Evaluate model performance R-value $\eta$ and MSE $\epsilon$ on the validation set
 \If{$\eta > \hat{\eta}$ and $\epsilon < \hat{\epsilon}$}
  \State $\hat{\eta} \coloneqq \eta$
  \State $\hat{\epsilon} \coloneqq \epsilon$
  \State Generate pseudo BMDs for unlabeled images
 \EndIf
 \State Fine-tune model on labeled images and unlabeled images with pseudo BMDs
 \State $e \coloneqq e+1$\
\EndWhile
\end{algorithmic}
\end{algorithm}

\section{Experiment Results}

\textbf{Dataset and evaluation metrics.}
We collected 1,090 hip X-ray images with associated DEXA-measured BMD values from 819 patients. The X-ray images are taken within six months of the BMD measurement. We randomly split the images into training, validation, and test sets of 440, 150, and 500 images based on patient identities. For semi-supervised learning, 8,219 unlabeled hip X-ray images are collected. To extract hip ROIs around the femoral neck, we train a localization model with the deep adaptive graph (DAG) network (re-implemented from \cite{li2020structured}) using about 100 images with manually annotated anatomical landmarks. We adopt two metrics, Pearson correlation coefficient (R-value) and Root Mean Square Error (RMSE), to evaluate the proposed method and all compared methods.

\textbf{Implementation details.}
The ROIs are resized to $512 \times 512$ pixels as model input. Random affine transformations, color jittering, and horizontal flipping are applied to each ROI during training. We adopt VGG-11 with batch normalization~\cite{ioffe2015batch} and squeeze-and-excitation (SE) layer~\cite{hu2018squeeze} as the backbone, since it outperforms other VGG networks~\cite{simonyan2014very} and ResNets~\cite{he2016deep} as demonstrated later.
The margin for adaptive triplet loss is set to 0.5. We use Adam optimizer with a learning rate of $10^{-4}$ and weight decay of $4\times10^{-4}$ to train the network on labeled images for 200 epochs. The learning rate is decayed to $10^{-5}$ after 100 epochs. For semi-supervised learning, we fix the learning rate at $10^{-5}$ to fine-tune for another 100 epochs. After each training/fine-tuning epoch, we evaluate the model on the validation set and select the one with the highest Pearson correlation coefficient for testing. All models are implemented using PyTorch 1.7.1 and trained on a workstation with an Intel(R) Xeon(R) CPU, 128G RAM, and a 12G NVIDIA Titan V GPU. We set the batch size to 16. The adaptive triplet loss is calculated for triplets randomly constructed within each batch. Specifically, each sample is used as an anchor with two other samples randomly selected as the near/far samples depending on their BMD distances to the anchor.

{\bf Backbone selection.} Here we study how different backbones affect the baseline performance without adaptive triplet loss or self-training. The compared backbones include VGG-11, VGG-13, VGG-16, ResNet-18, ResNet-34, and ResNet-50. As shown in Table~\ref{tab:baseline_comparison}, VGG-11 achieves the best R-value of 0.8520 and RMSE of 0.0831. The lower performance of other VGG networks and ResNets may be attributed to overfitting from more learnable parameters. In light of this, we conjecture there might be a more appropriate network backbone for BMD regression, which is outside this paper's scope. Therefore, we adopt VGG-11 with batch normalization and SE layer as our network backbone.

\setlength{\tabcolsep}{2mm}
\begin{table}[t]
    \centering
    \caption{Comparison of baseline methods using different backbones.}
    \vspace{-1em}
    \begin{tabular}{ccccccc}
    \toprule
    Backbone & VGG-11 & VGG-13 & VGG-16 & ResNet-18 & ResNet-34 & ResNet-50 \\
    \hline
    R-value & \textbf{0.8520} & 0.8501 & 0.8335 & 0.8398 & 0.8445 & 0.8448 \\
    RMSE & \textbf{0.0831} & 0.0855 & 0.1158 & 0.0883 & 0.0946 & 0.1047 \\
    \bottomrule
    \end{tabular}
    \label{tab:baseline_comparison}
    \vspace{-2em}
\end{table}

{\bf Quantitative comparison results.} We compare our method with three existing semi-supervised learning (SSL) methods: $\rm{\Pi}$-model~\cite{laine2016temporal}, temporal ensembling~\cite{laine2016temporal}, and mean teacher~\cite{tarvainen2017mean}. Regression MSE loss between predicted and GT BMDs is used on labeled images for all semi-supervised learning methods. $\rm{\Pi}$-model is trained to encourage consistent network output between two augmentations of the same input image.
Temporal ensembling produces pseudo labels via calculating the exponential moving average of predictions after every training epoch. Pseudo labels are then combined with labeled images to train the model. Instead of directly ensembling predictions, mean teacher uses an exponential moving average of model weights to produce pseudo labels for unlabeled images. For the proposed method and compared SSL methods, all models are fine-tuned from the weights pre-trained on labeled images.

\setlength{\tabcolsep}{4mm}
\begin{table}[t]
    \centering
    \caption{Comparison with semi-supervised learning methods. (Temp. Ensemble: temporal ensembling)}
    \vspace{-1em}
    \begin{tabular}{ccccc}
    \toprule
    Method  & $\rm{\Pi}$-model & Temp. Ensemble & Mean Teacher & Proposed \\
    \hline
    R-value & 0.8637 & 0.8722 & 0.8600 & \textbf{0.8805} \\
    RMSE & 0.0828 & 0.0832 & 0.0817 & \textbf{0.0758} \\
    \bottomrule
    \end{tabular}
    \label{tab:ssl_comparison}
    \vspace{-1em}
\end{table}

\begin{figure}[t]
    \centering
    \includegraphics[width=0.6\linewidth]{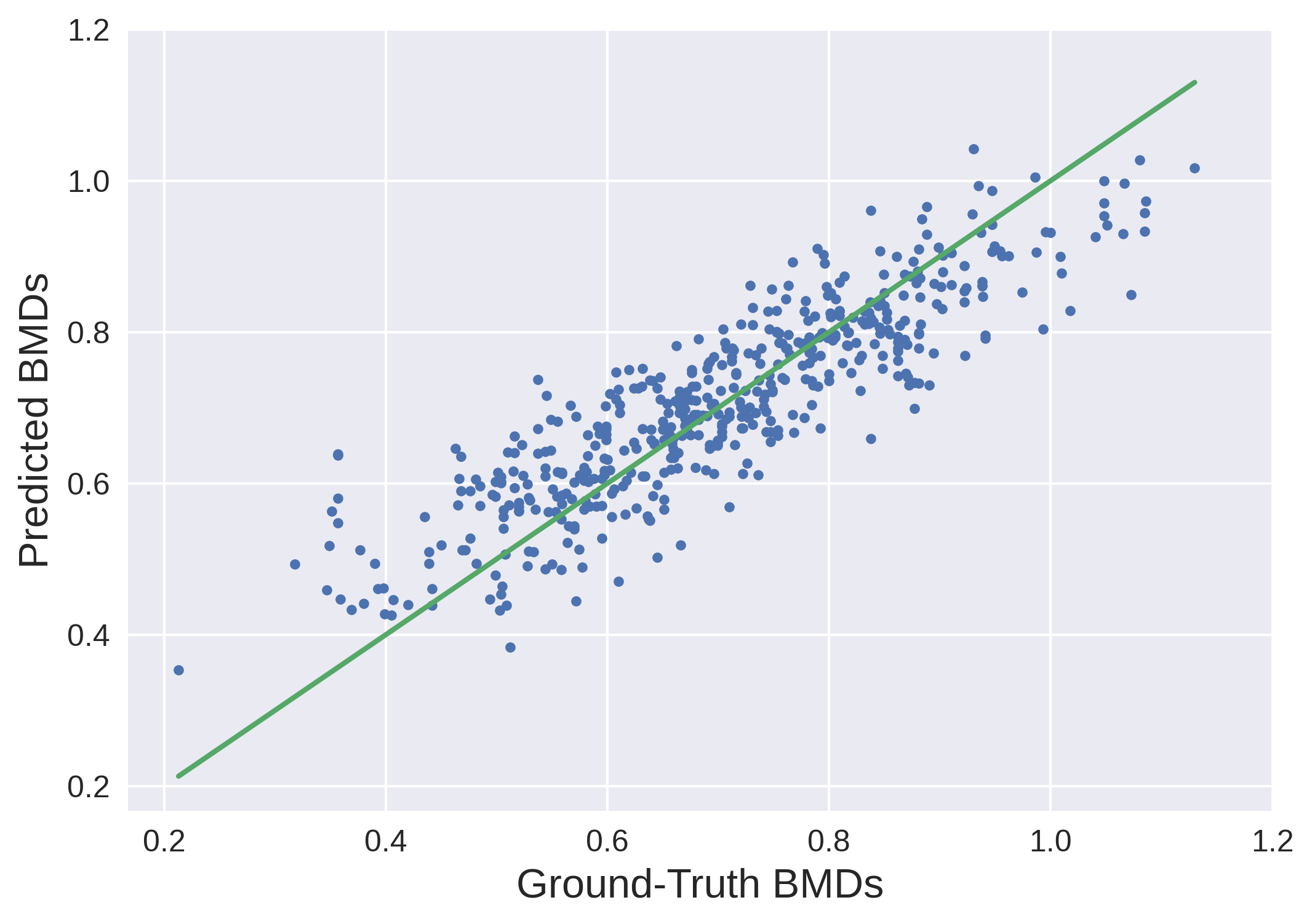}
    \vspace{-1em}
    \caption{Predicted BMDs against ground-truth BMDs on the test set.}
    \label{fig:bmd_plot}
    \vspace{-2em}
\end{figure}

As shown in Table~\ref{tab:ssl_comparison}, the proposed method achieves the best R-value of 0.8805 and RMSE of 0.0758. $\rm{\Pi}$-model outperforms the baseline by enforcing output consistency as a regularization. While both temporal ensembling and mean teacher obtain improvements with the additional pseudo label supervision, averaging labels or weights can accumulate more errors over time. In contrast, our method only updates pseudo labels when the model performs better on the validation set, proving to be an effective strategy.
Figure~\ref{fig:bmd_plot} shows the predicted BMDs of our model against GT BMDs. Overall, the model has a larger prediction error for lower or higher BMD cases. This situation is expected because lower/higher BMD cases are less common, and the model tends to predict moderate values.

{\bf Ablation study.} To assess the importance of various components in our proposed method, we conduct experiments to remove some components or vary some hyper-parameters while keeping others fixed. Firstly, we analyze the proposed adaptive triplet loss (ATL) by comparing it to its counterpart without the adaptive coefficient in Eq.~\ref{eq:adaptive_coefficient} and varying the preset margin. Results in Table~\ref{tab:ablation_triplet} show that the non-adaptive counterpart deteriorates the model's regression accuracy, which proves the necessity of adaptive coefficient. Since BMD differences vary for different triplets, it is unreasonable to use a fixed margin to uniformly separate samples with dissimilar BMDs. We also observe that using ATL achieves higher R-values than the baseline regardless of the margin value. Specifically, $m=0.5$ produces the best R-value of 0.8670 and RMSE of 0.0806. It demonstrates the effectiveness of our proposed adaptive triplet loss.

\setlength{\tabcolsep}{4mm}
\begin{table}[t]
\centering
\caption{Ablation study of adaptive triplet loss.}
\vspace{-1em}
\begin{tabular}{ccccc}
\toprule
\multirow{2}{*}{Margin}  & \multicolumn{2}{c}{Triplet Loss} & \multicolumn{2}{c}{ATL} \\ 
                           \cmidrule(lr){2-3}                 \cmidrule(lr){4-5}
              & R-value & RMSE                  & R-value & RMSE  \\ \midrule
0.1           & 0.8515  & 0.0917                & 0.8563  & 0.1013\\
0.3           & 0.8459  & 0.1000                & 0.8657  & 0.0814 \\
0.5           & 0.8538  & 0.0866                & \textbf{0.8670}  & \textbf{0.0806} \\
0.7           & 0.8549  & 0.0823                & 0.8565  & 0.0836 \\
1.0           & 0.8522  & 0.0829                & 0.8524  & 0.1215 \\
\bottomrule
\end{tabular}
\label{tab:ablation_triplet}
\end{table}

To better understand the benefits of our proposed semi-supervised self-training algorithm, we compare it with a straightforward semi-supervised learning (SSL) approach. We use the pre-trained model to predict pseudo BMDs for unlabeled images and combine them with labeled ones for further fine-tuning. We apply this approach to fine-tune models pre-trained using only MSE loss and the combination of MSE and ATL losses, denoted as SSL and SSL+ATL, respectively. Finally, we also analyze the contribution of the consistency loss in Eq.~\ref{eq:consistency_loss} by removing it during the self-training stage. The comparison results are shown in Table~\ref{tab:ablation_proposed_method}.

\setlength{\tabcolsep}{4mm}
\begin{table}[t]
    \centering
    \caption{Ablation study of adaptive triplet loss and self-training algorithm.}
    \vspace{-1em}
    \begin{tabular}{ccc}
    \toprule
    Method & R-value & RMSE \\
    \cmidrule(lr){1-1}  \cmidrule(lr){2-3}
    Baseline                 & 0.8520 & 0.0831 \\
    Baseline + ATL           & 0.8670 & 0.0806 \\ \midrule
    SSL                      & 0.8605 & 0.0809 \\
    SSL + ATL                & 0.8772 & 0.0767 \\ \midrule
    Proposed w/o Consistency & 0.8776 & 0.0761 \\
    Proposed                 & \textbf{0.8805} & \textbf{0.0758} \\
    \bottomrule
    \end{tabular}
    \label{tab:ablation_proposed_method}
    \vspace{-2em}
\end{table}

Employing the straightforward SSL strategy is effective as expected. It increases the baseline R-value to 0.8605 and reduces the RMSE to 0.0809. The model pre-trained using MSE and ATL losses is also improved. The results prove the effectiveness of using pseudo labels of unlabeled images. Our proposed self-training algorithm can obtain further improvement by updating pseudo labels during fine-tuning. However, the results show that the improvement becomes marginal from 0.8772 to 0.8776 in R-value without the consistency loss, even if pseudo labels are updated a few more times. The consistency loss can regularize model training by encouraging consistent output and features. Without it, the model is prone to overfitting to inaccurate pseudo labels and deteriorates. Combining ATL, self-training algorithm, and consistency loss, our proposed method achieves the best R-value of 0.8805 and RMSE of 0.0758. Compared to the baseline, the R-value is improved by 3.35\% and the RMSE is reduced by 8.78\%.

\section{Conclusion}
In this work, we investigate the feasibility of obtaining bone mineral density (BMD) from hip X-ray images instead of DEXA measurement. It is highly desirable for various practical reasons. We employ a convolutional neural network to estimate BMDs from preprocessed hip ROIs. Besides the MSE loss, a novel adaptive triplet loss is proposed to train the network on hip X-ray images with paired ground-truth BMDs. We further present a semi-supervised self-training algorithm to improve our model using large-scale unlabeled hip X-ray images. Experiment results on an in-house dataset of 1,090 images from 819 patients show that our method effectively achieves a high Pearson correlation coefficient of 0.8805. It implies the strong feasibility of X-ray based BMD estimation, which can provide potential opportunistic osteoporosis screening with more accessibility and at reduced cost.

%
%
%
\bibliographystyle{splncs04}
\bibliography{miccai2021}
\end{document}